\documentclass[preprint,showpacs,preprintnumbers,amsmath,amssymb]{revtex4}

\usepackage{graphicx}
\usepackage{epsfig}

\begin{document}

\newcommand{\bc}{\ensuremath{\mathbf{c}}}
\newcommand{\bj}{\ensuremath{\mathbf{j}}}
\newcommand{\normSt}{\ensuremath{{N}}}
\newcommand{\Kn}{\ensuremath{{\rm Kn}}}

\title{Hydrodynamics beyond Navier-Stokes:\\ Exact solutions to the lattice Boltzmann hierarchy}

\author{S. Ansumali}
\affiliation{School of Chemical and Biomedical Engineering,
Nanyang  Technological University, 639798 Singapore}

\author{ I. V. Karlin\footnote{Corresponding author}}
\affiliation {Aerothermochemistry and Combustion Systems Lab, ETH
Zurich, 8092 Zurich, Switzerland}

\author{S. Arcidiacono}
\affiliation {Paul Scherrer Institute, Combustion Research, 5232
Villigen PSI, Switzerland}

\author{A. Abbas}
\affiliation{School of Chemical and Biomedical Engineering,
Nanyang Technological University, 639798 Singapore, Singapore}

\author{ N. I. Prasianakis}
\affiliation {Aerothermochemistry and Combustion Systems Lab, ETH
Zurich, CH-8092 Zurich, Switzerland}

\begin{abstract}
 Exact solution to the hierarchy of nonlinear lattice Boltzmann (LB) kinetic
equations in the stationary planar Couette flow is found at
non-vanishing Knudsen numbers. A new method of solving LB kinetic
equations which combines the method of moments with boundary
conditions for populations enables to derive closed-form solutions
for all higher-order moments. Convergence of results  suggests
that the LB hierarchy with larger velocity sets is the novel way
to approximate kinetic theory.
\end{abstract}
 \maketitle


Emerging field of fluid dynamics at a micrometer scale becomes
increasingly important due to fundamental engineering issues of
micro-electromechanical systems \cite{Karniadakis2}. Recently,
much attention was attracted by the use the lattice Boltzmann (LB)
models for simulation of microflows by a number of groups
\cite{AK4,SucciPRL02,KWOK,Niu04,AK7,Sofonea05,SucciPRL06}. By now,
it is understood that lattice Boltzmann models form a well-defined
hierarchy based on discrete velocity sets with velocities defined
as zeroes of Hermite polynomials \cite{AK5} or rational-number
approximations thereof \cite{ShyamKarlinPRL06}. The LB hierarchy
constitutes a novel approximation of the Boltzmann equation and
has to be considered as an alternative to more standard approaches
such as higher-order hydrodynamics (Burnett or super-Burnett) or
Grad's moment systems (for a review see, e. g. \cite{GKbook}). One
salient feature of the LB hierarchy, which is crucial to the
present study and eventually to any realistic application, is that
it is naturally equipped with relevant boundary conditions derived
from Maxwell-Boltzmann's theory \cite{AK4}.

Agreement between the LB simulations and kinetic theory
\cite{AK4}, hydrodynamics with slip boundary conditions
\cite{Sofonea05}, and molecular dynamics \cite{SucciPRL06} was
reported. However, most of these numerical works rely on
simulation with finite accuracy while the crucial question whether
or not the kinetic equations underpinning the LBM method are valid
physical models of microflow remains unanswered. Therefore, it is
not surprising to read comments claiming, for example, that the
slip flow in the LBM is due to discretization errors rather than a
physical effect,
(\cite{LUOCOMMENT} and references therein).
Thus, validity of LBM cannot be addressed unless a comparison to
representative exact solutions is performed. It is needless to say
that exact solutions to nonlinear kinetic equations in realistic
geometries are very rare.


In this Letter, we show that the LB hierarchy of kinetic models
admits a much more accurate treatment. In particular, we find
closed-form analytical solutions to nonlinear kinetic equations of
the LB hierarchy in the stationary planar Couette flow.
Not only the slip velocity, but also the  shear stress and the
normal stress difference are evaluated in a closed form.
Comparison to the kinetic theory demonstrates convergence of
approximations with the increase of the number of velocities. In
the nonlinear domain, even the first member of the LB hierarchy
predicts nontrivial normal stress which is confirmed with a more
microscopic direct simulation Monte Carlo (DSMC) method
\cite{Bird}. The accurate results obtained herein strongly suggest
that the LB hierarchy should be considered as a novel general tool
of kinetic theory rather than a plain solver for hydrodynamics.


Kinetic equations studied in this paper are  two-dimensional
isothermal discrete velocity models with the Bhatnagar-Gross-Krook
(BGK) nonlinear collision integral (for a derivation of these
models from the Boltzmann-BGK equation see, e. g.,
\cite{AK5,ShyamKarlinPRL06}),
\begin{equation}
\label{LBM}
\partial_t f_i+ c_{i\alpha}\partial_{\alpha}f_i =
-\frac{1}{\tau} \left(f_i- f_i^{\rm eq}  \right),
\end{equation}
where summation convention is applied, $\tau$ is the relaxation
time, and the equilibrium distribution is
\begin{align}
\label{TQA}
\begin{split}
  f_i^{\rm eq}(\rho,j_x,j_y) =w_i\left(\rho  + \frac{j_{\alpha}c_{i\, \alpha}}{c_{\rm s}^2}
  +
\frac{j_{\alpha} \,j_{\beta }}{ 2\rho c_{\rm s}^4}
    \left(c_{i\alpha} c_{i\beta} -  c_{\rm s}^2  \delta_{\alpha \beta }\right)
   \right).
\end{split}
\end{align}
Here $c_{\rm s}= \sqrt{(k_{\rm B}T_0)/m}$ is the speed of sound,
$T_0$ is the reference temperature.

The first member of the LB hierarchy is the so-called $D2Q9$ model
where the discrete velocities $c_{i\alpha}$, $i=0,\dots, 8$, and
the weights $w_i$ are
\begin{align}
\label{dvSt}
\begin{split}
c_x &= \sqrt{3}c_{\rm s}\left\{0, 1, 0, -1, 0, +1,-1,-1,+1\right\}\\
c_y &= \sqrt{3}c_{\rm s}\left\{0, 0, 1,  0, -1, +1,+1,-1,-1\right\}\\
w &=(1/36) \left\{16, 4, 4,  4, 4, 1,1,1,1\right\}.
\end{split}
\end{align}
The model (\ref{LBM}) conserves locally the density
$\rho=\sum_{i=0}^{8}f_i$ and the momentum density
$j_{\alpha}=\sum_{i=0}^{8} c_{i\alpha}f_i$ but not the energy. In
the hydrodynamic limit, it recovers the Navier-Stokes equations
with the kinematic viscosity
 $\nu = \tau c_{\rm s}^2 $.

We consider the planar Couette flow, where a fluid is enclosed
between two parallel plates separated by a distance $L$. The
bottom plate at $y=-L/2$ moves with the velocity $U_1$ and top
plate at $y=L/2$ moves with the velocity $U_2$ (see Fig.\
\ref{Fig1}). Let us introduce the mean free path $l=\sqrt{3}\tau
c_{\rm s}$ and the Knudsen number $\Kn = l/L$. The solution for
the $x$-velocity of the $D2Q9$ model derived below reads:
\begin{center}
\begin{figure}[htp]
 \includegraphics[scale=0.6]{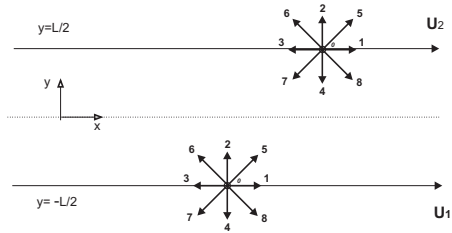}
 \caption{Couette flow geometry. Discrete velocities of the $D2Q9$ model
 at the bottom and the top plates are indicated to explain boundary
 conditions.\label{Fig1}}
 \end{figure}
 \end{center}
%
\begin{equation}
 u_x=\frac{1}{\Theta_{9}}
\left(\frac{y}{L}\right)\Delta U+U,\label{VelocityResult}
\end{equation}
where \begin{equation}
\Theta_{9}=
 1+ 2\Kn,
 \label{Slope9}
 \end{equation}
and
 where $\Delta U=U_2-U_1$ is the relative velocity of the plates,
 and $U=(U_1+U_2)/2$ is the centerline velocity.
Solution (\ref{VelocityResult}) is Galilean invariant: if a
constant velocity $C$ is added to both the plates,
$U_{1,2}'=U_{1,2}+C$ then the velocity (\ref{VelocityResult})
transforms as $u_x'=u_x+C$. The slip velocity at the plates,
$u_x(\pm L/2)$, features the expected behavior from a linear
increase with $\Kn$ at small $\Kn$ to a plug flow at
$\Kn\to\infty$ where $u_x$ becomes position-independent.  Recent
careful numerical study of the $D2Q9$ model by Sofonea and Sekerka
\cite{Sofonea05} revealed the same result (\ref{VelocityResult})
and (\ref{Slope9}).

\begin{figure}
\centering
\begin{tabular}{l}
\epsfig{file=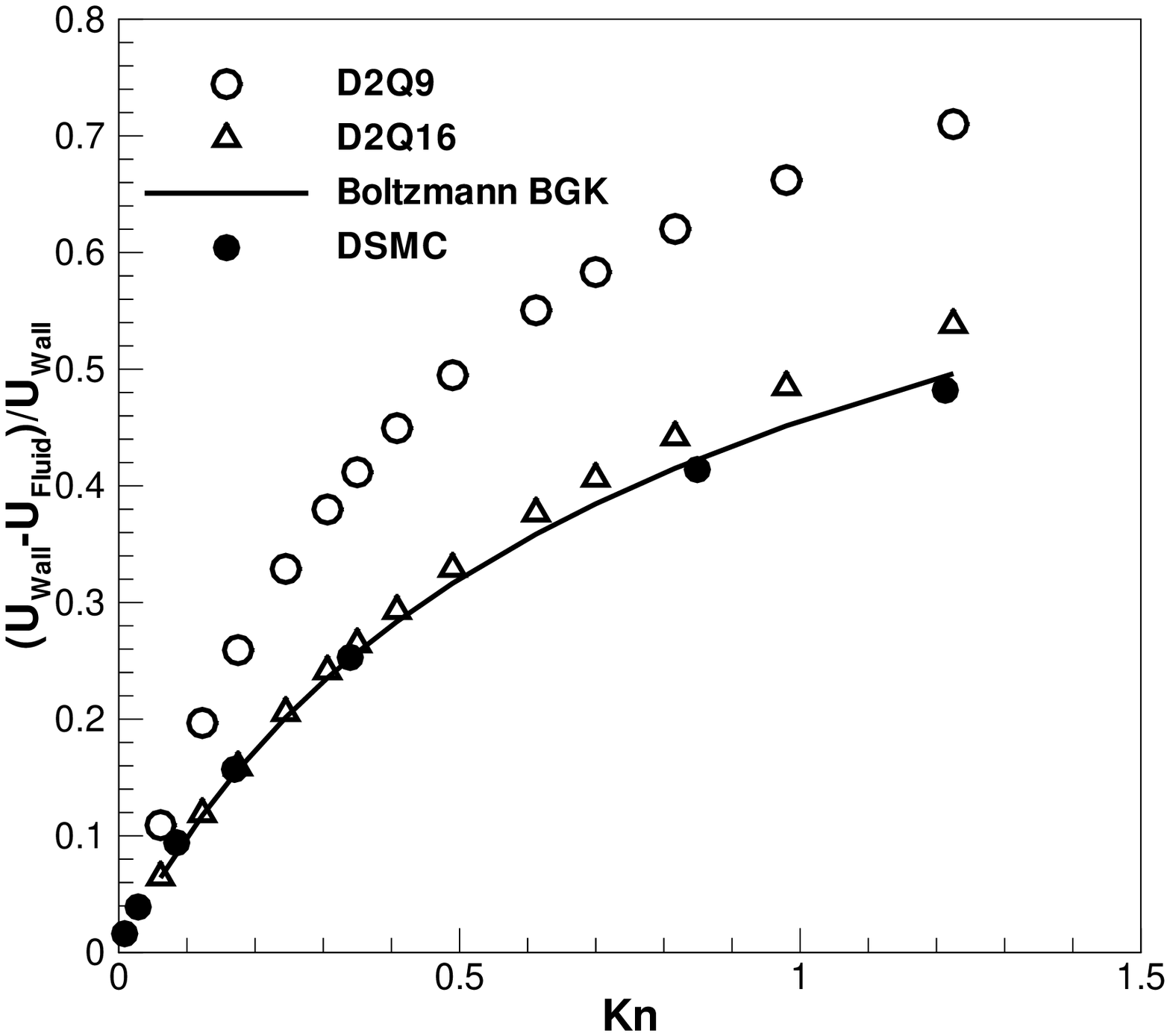,width=0.5\linewidth,clip=}
\epsfig{file=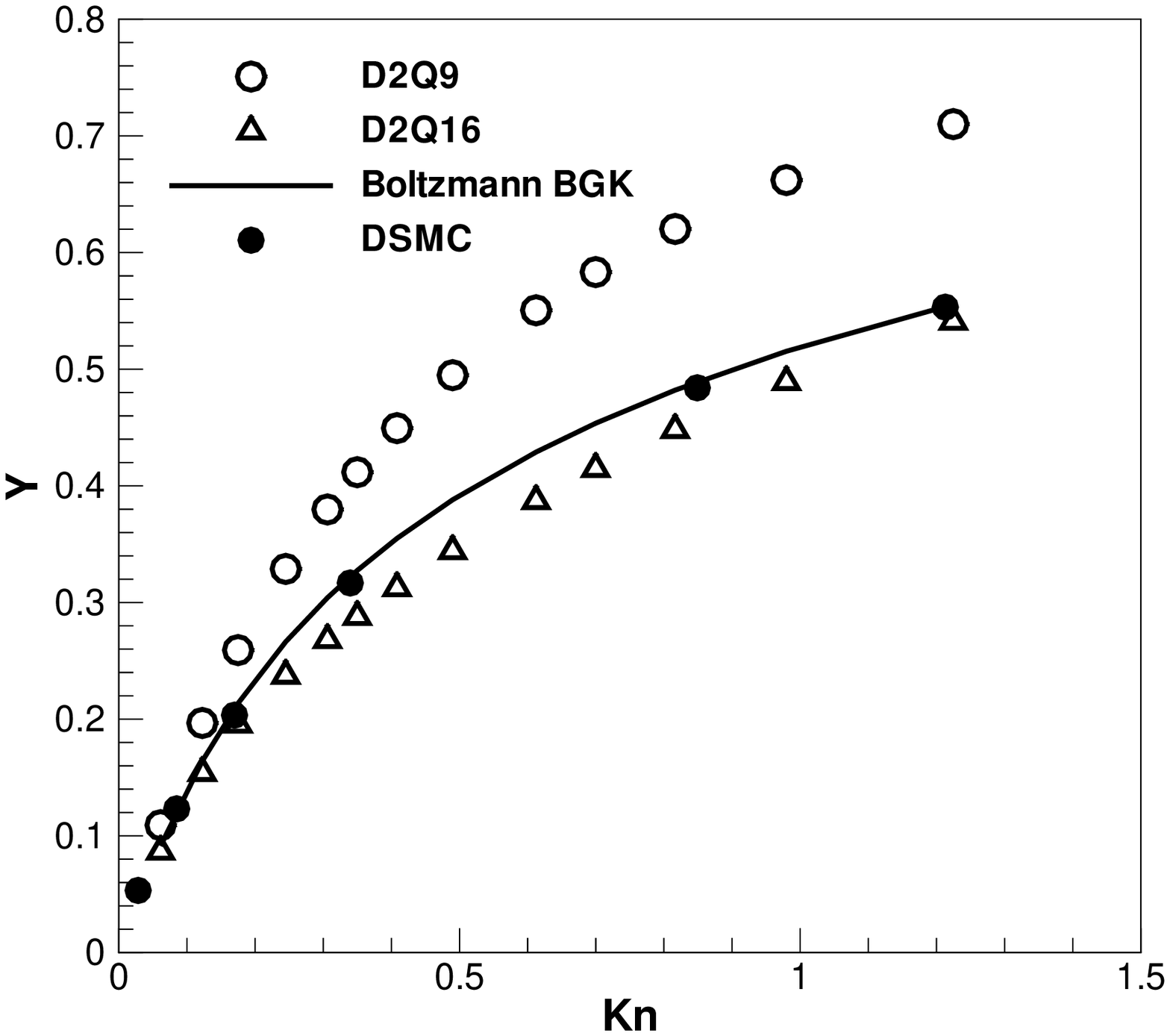,width=0.5\linewidth,clip=}
\end{tabular}
\caption{Comparison of the LB hierarchy with the linearized
Boltzmann-BGK model \cite{Willis62} and DSMC simulation. Left:
Slip velocity at the wall as a function of Knudsen number. Right:
Slope of the velocity profile at the centerline. Plotted is the
deviation from the Navier-Stokes prediction, $Y=\Delta
U^{-1}(du_x/dy)\big|_{y=0}-1$.} \label{Fig2}
\end{figure}

The next member of the LB hierarchy is the model based on the
roots of fourth-order Hermite polynomial $\{\pm a,\pm b\}$, where
$a=\sqrt{3-\sqrt{6}}$ and $b=\sqrt{3+\sqrt{6}}$. In two
dimensions, the discrete velocities are all possible tensor
products of the two copies of the four-sets $\{\pm a,\pm b\}$. For
this $D2Q16$ model \cite{AK5,ShyamKarlinPRL06}, the solution for
the velocity profile is found to be:
\begin{equation}
u_x =\frac{1}{Z_{16}} \sinh{\left(\frac{y}{ \Kn L}\right)}\Delta U
+ \frac{1}{\Theta_{16}}\left(\frac{ y}{ L}\right)\Delta U +U,
\label{VelocityResult2}
\end{equation}
where
\begin{eqnarray}
\Theta_{16}&=&1+2\Kn\left(\frac{
   2\cosh \left(\frac{1}{2 \, \Kn }\right)+
   \mu\sinh \left(\frac{1}{2
   \text{Kn}}\right)}{\mu
   \cosh \left(\frac{1}{2 \text{Kn}}\right)+  2\sqrt{3} \, \sinh \left(\frac{1}{2
   \text{Kn}}\right)}\right),\label{Slope16}\\
Z_{16}&=&
    \frac{\mu }{4\Kn}   \left(
   \left(4\Kn+\mu\right)\cosh \left(\frac{1}{2 \text{Kn}}\right)+2\left(\mu\Kn
   + \sqrt{3}
    \right)\sinh \left(\frac{1}{2 \text{Kn}}\right)\right),
\end{eqnarray}
and $\mu=a+b\approx 3.076$. The difference between
(\ref{VelocityResult}) and (\ref{VelocityResult2}) is that the
latter predicts the boundary Knudsen layer (first term in
(\ref{VelocityResult2})) in a qualitative agreement with kinetic
theory \cite{Cerci}. On the quantitative side, our analytical
results can be immediately compared with the classical study of
the linearized Boltzmann-BGK equation by Willis \cite{Willis62}
(see Fig.\ \ref{Fig2}, where also the results of the DSMC
simulation are reported; note that data in Fig.\ \ref{Fig2} is
parameterized with the Knudsen number according to a relation,
$\Kn=\sqrt{\frac{3}{2}}\alpha^{-1}$, where
$\alpha=\frac{L}{\tau}\sqrt{\frac{m}{2k_{\rm B}T_0}}$ is a
parameter used in Table I of Ref.\ \cite{Willis62}). While the
simplest $D2Q9$ model predicts well a slip-flow solution, it fails
in the transient regime ($\Kn\gtrsim 0.1$), in agreement with
numerical studies \cite{AK4,Guo06bs}. However, already the $D2Q16$
model considerably improves the situation.
The strong
pattern of convergence with increasing the number of velocities in
the LB hierarchy is clearly there.


We shall now proceed with major steps of derivation for the $D2Q9$
model. Firstly, we write the kinetic equation for nine populations
(\ref{LBM}) in a form of a moment system for nine moments which we
choose as follows: Three locally conserved fields, $\rho$, $j_x$,
$j_y$; three independent components of the pressure tensor,
$P_{\alpha \beta} = \sum_{i=0}^{8} \, f_i \, c_{i\alpha} \, c_{i
\beta}$, which we choose as the trace $P=P_{xx}+P_{yy}$,
$N=P_{xx}-P_{yy}$ (normal stress difference), and $P_{xy}$; two
components of the energy flux, $q_{\alpha} =\sum_{i=0}^{8} f_i c_{
\alpha i} c_{i}^2 $, and a scalar fourth-order moment, $ \psi
=R_{yyyy}+R_{xxxx}- 2 R_{xxyy}$, where $R_{\alpha \beta \gamma
\theta}=\sum_{i=0}^{8} f_i c_{i\alpha} c_{ i\beta} c_{ i\gamma}
c_{ i\theta }$. The resulting moment system includes three balance
equations for the locally conserved fields,
\begin{eqnarray}
 \partial_t \rho +  \, \partial_x j_x +   \partial_y j_y &=& 0,
 \nonumber\\
 \partial_t j_x + \frac{1}{2}\partial_x(P+ N)+  \partial_y P_{x y} &=& 0, \nonumber\\
 \partial_t j_y + \partial_x P_{xy} + \frac{1}{2}\partial_y (P- N) &=&
 0, \label{ConsEqn}
\end{eqnarray}
three equations for the components of the pressure tensor,
\begin{eqnarray}
 \partial_t P_{xy} \!+\!\partial_x (q_y  -  3c_{\rm s}^2j_y) \!+\!\partial_y(q_x - 3c_{\rm s}^2j_x)= \frac{1}{\tau}
(\frac{j_x j_y }{\rho}-P_{xy}), \nonumber\\
 \partial_t \,\normSt \!+\!  \partial_x\,(6c_{\rm s}^2j_x - q_x)\! +\!\partial_y(q_y  -  6c_{\rm s}^2j_y) =
\frac{1}{\tau}\,(\frac{j_x^2 }{\rho} -\frac{j_y^2 }{\rho} -\normSt), \nonumber\\
\partial_t P \!+\! \partial_x q_x \!+\! \partial_y q_y  =\frac{1}{\tau}\,(2\rho \, c_s^2 +\frac{j^2 }{\rho}   -
P), \label{PtensEv}
\end{eqnarray}
two equations for the components of the energy flux,
\begin{eqnarray}
\partial_t q_x +\partial_x [3c_{\rm s}^2\,(P  +  \frac{1}{2}\normSt) -
\frac{1}{2}\psi] +   6c_{\rm s}^2\partial_y P_{xy} &=&
    \frac{1}{\tau}\,( 4\, c_{\rm s}^2 \, j_{x}   - q_{x}),\nonumber\\
\partial_t q_y +  6c_{\rm s}^2\partial_x P_{xy}+  \partial_y [3c_{\rm s}^2\,
(P  - \frac{1}{2}\normSt) - \frac{1}{2}\psi]&=&
    \frac{1}{\tau}( 4\, c_{\rm s}^2 j_{y}   -
    q_{y}),\label{qevol}
 \end{eqnarray}
and, finally, the equation for the fourth-order moment $ \psi$,
\begin{equation}
 \partial_t \psi + 3\, c_{\rm s}^2\,
\partial_x \,( 6c_{\rm s}^2j_x-q_{x}) +
3\, c_{\rm s}^2 \,  \partial_y(6c_{\rm s}^2j_y-q_{y}) =
\frac{1}{\tau}( 4\rho c_{\rm s}^4 + c_{\rm s}^2\frac{j^2}{\rho} -
\psi). \label{fourevol}
\end{equation}

Secondly, we find steady state solution to the moment system under
two conditions: (i) Unidirectional flow: As the plates extend to
infinity in the $x$ direction, we can expect that the steady state
solution will be independent of $x$, and (ii) Impermeable plates:
The normal mass flux equals zero at the walls. For a
unidirectional stationary flow, balance equations (\ref{ConsEqn})
read: $\partial_y j_y =0$, $\partial_y P_{xy} =0$, and
$\partial_y(P- \normSt)= 0$, whereupon,
%
using condition (ii), we get
$j_y = 0$,
$P_{xy}=P_{xy}^{\rm neq}$,
$P= \normSt + P_0$,
where
%
%
integration constants $P_{xy}^{\rm neq}$ and $P_0$ will be
determined below (superscript `neq' is added to emphasize that
only the non-equilibrium part is nontrivial for $P_{xy}$).
Furthermore, equation for the pressure tensor (\ref{PtensEv})
reads
\begin{eqnarray}
\label{Step31}
\partial_y (q_x - 3\, c_{\rm s}^2 j_x) &=&- \frac{1}{\tau} P_{xy}^{\rm neq},\\
 \label{Step32}
\partial_y q_{y}   &=& \frac{1}{\tau}\, ( \frac{j_x^2 }{{\rho}}  -\normSt),\\
\label{Step33}
 \partial_y q_y  &=&\frac{1}{\tau}\,(2\, {\rho} \, c_{\rm s}^2 +\frac{j_x^2 }{{\rho}}   - \normSt -
 P_0),
\end{eqnarray}
From (\ref{Step32}) and (\ref{Step33}) it follows
%
 $P_{0} = 2{\rho}c_{\rm s}^2$.
Thus, the stationary density is constant. Equation for energy flux
(\ref{qevol}) simplifies as:
\begin{eqnarray}
\label{Step51}
  q_x &=& 4c_{\rm s}^2 \, j_x   , \\
 \partial_y [ 3\, c_{\rm s}^2\, (P -  \frac{1}{2}\normSt) -
\frac{1}{2}\psi]
 &=&-\frac{1}{\tau}q_y.\label{Step52}
\end{eqnarray}
Substituting (\ref{Step51}) into (\ref{Step31}), and integrating
the resulting differential equation for $j_x$, we obtain the
result for the nontrivial velocity component $u_x=j_x/{\rho}$,
\begin{equation}
\label{VelExp}
 u_x(y)
= -\frac{\pi}{\tau\, c_{\rm s}^2}\left(y + \tau\, V\right),\\
\end{equation}
where $V$ is constant of integration, and where we have introduced
$\pi=P^{\rm neq}_{xy}/\rho$.
Thus, we find the solution for the velocity (up to two constants,
$\pi$ and $V$) {\it before} higher-order moments are addressed. We
note in passing that it is precisely the relation (\ref{Step51})
pertinent to the low-symmetry $D2Q9$ model (the energy flux is
proportional to the momentum flux) which precludes the development
of the boundary Knudsen layer. This constraint is removed in the
more symmetric $D2Q16$ model and in any higher-order member of the
LB hierarchy.

The stationary equation for fourth order moment $\psi$
(\ref{fourevol}), together with (\ref{Step32}), gives
 $\psi = 4c_{\rm s}^2 \frac{j_x^2 }{{\rho}}- 3c_{\rm s}^2\normSt+ 4{\rho} c_{\rm s}^4$.
Finally, from (\ref{Step52}) and  (\ref{Step32}), we get
\begin{equation}
 \partial_y q_{y} ^{\rm neq} =- \frac{\normSt^{\rm neq}}{\tau},\
 \partial_y\normSt^{\rm neq}  =-\frac{q_y^{\rm neq}}{3\tau c_s^2}-\frac{{\rho}}{3}\partial_y
(u_x^2). \label{ODE}
\end{equation}
The ordinary differential  equations (\ref{ODE}) can be integrated
explicitly since the velocity $u_x(y)$ is available from
(\ref{VelExp}). Denoting $\varphi(y)=\exp{(y/\sqrt{3}\tau c_{\rm
s})}$, the result is
%
\begin{eqnarray}
 \sqrt{3} \, c_{\rm s}\, \normSt^{\rm neq} +
q_y^{\rm neq} &=&A \varphi(-y)
  -\frac{2{\rho}\pi^2}{\tau c_{\rm s}^2}
  \left[
  \,y + \tau(V-\sqrt{3}c_{\rm s})\left(1
 -\varphi(-y) \right)
\right],\nonumber\\
\sqrt{3} \, c_{\rm s}\, \normSt^{\rm neq} - q_y^{\rm neq}&=&B
\varphi(y)+\frac{2{ \rho}\pi^2}{\tau c_s^2}
  \left[
  y + \tau(V+ \sqrt{3}c_{\rm s})\, \left(1-
\varphi(y)\right) \right], \label{genexp1}
\end{eqnarray}
where $A$ and $B$ are constants of integration. Thus, the solution
to the stationary moment system depends on the four integration
constants, $\pi$, $V$, $A$ and $B$. In order to determine these,
we need to specify boundary conditions at the moving plates. Note
that this is precisely where the LB hierarchy differs from the
method of moments. It is well known that for moment methods, such
as Grad systems, it is not possible to provide self-consistent
boundary conditions for the moments. In our case, this is possible
because the boundary conditions for the LB equations are
formulated in terms of populations rather than in terms of
moments. Upon inverting the linear relation between the moments
and the populations, and using the solution for the moments
derived above, we obtain the stationary populations $f_i=f_i^{\rm
eq}+f_i^{\rm neq}$, where the stationary equilibrium part is given
by (\ref{TQA}) with $j_y=0$ and $j_x=\rho u_x$ (\ref{VelExp}),
while the non-equilibrium part has the form,
\begin{equation}
  f_i^{\rm neq}  = w_i\left(
  \frac{P_{xy}^{\rm neq}}{c_{\rm s}^4}
   c_{ix} c_{iy} \!+\!
\frac{q_{y}^{\rm neq}}{2c_{\rm s}^6}\left(c_{iy}c_i^2 \!-\! 4
c_{\rm s}^2 c_{iy} \right)\!+\!
 \frac{\normSt^{\rm neq}}{ 2 c_{\rm s}^6}
    \left(c_{ix}^2\!-\! c_{\rm s}^2 \right)c_{iy}^2
\right). \label{fneqsol}
\end{equation}

\begin{figure}
\centering
\begin{tabular}{l}
\epsfig{file=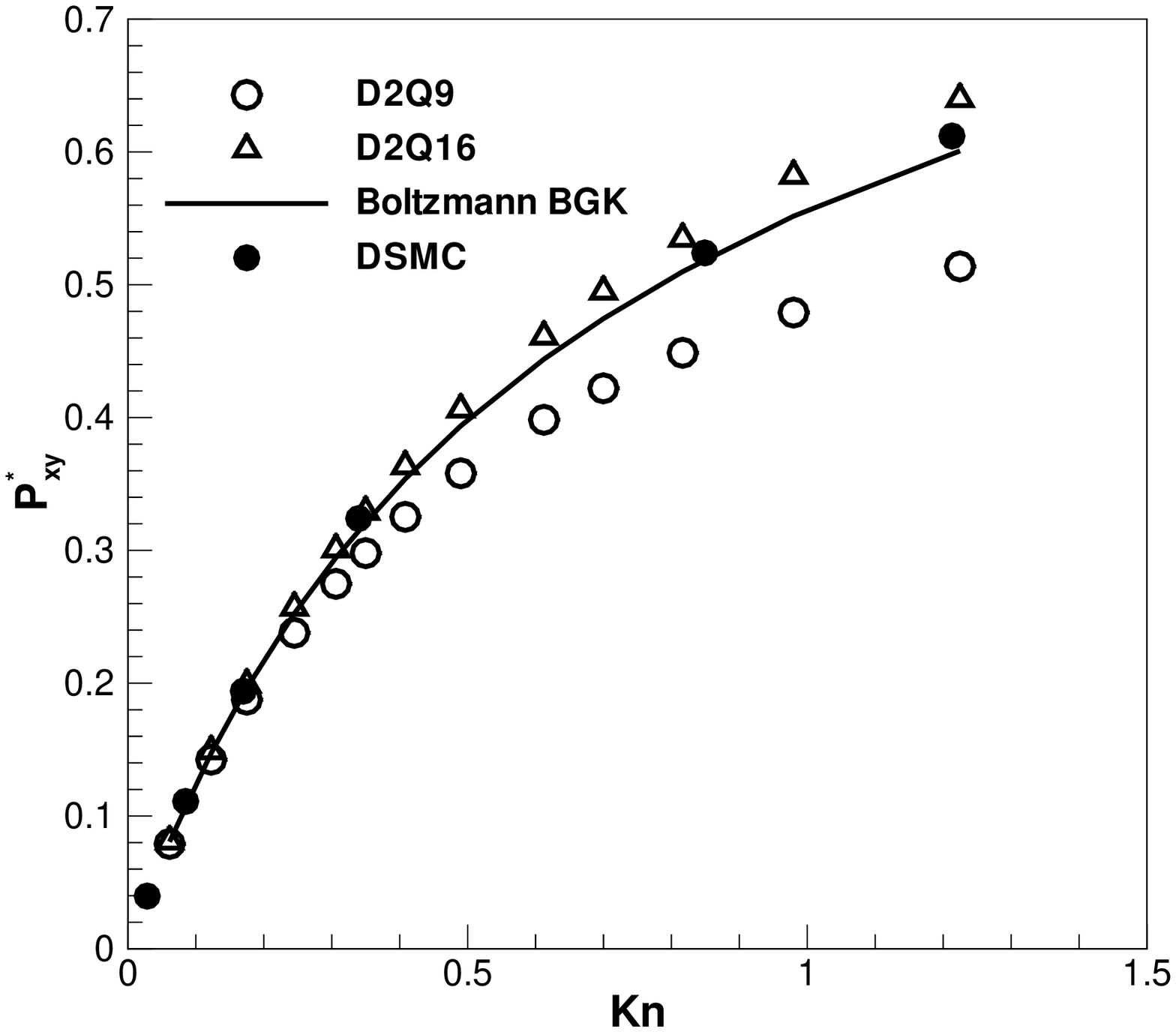,width=0.5\linewidth,clip=}
\epsfig{file=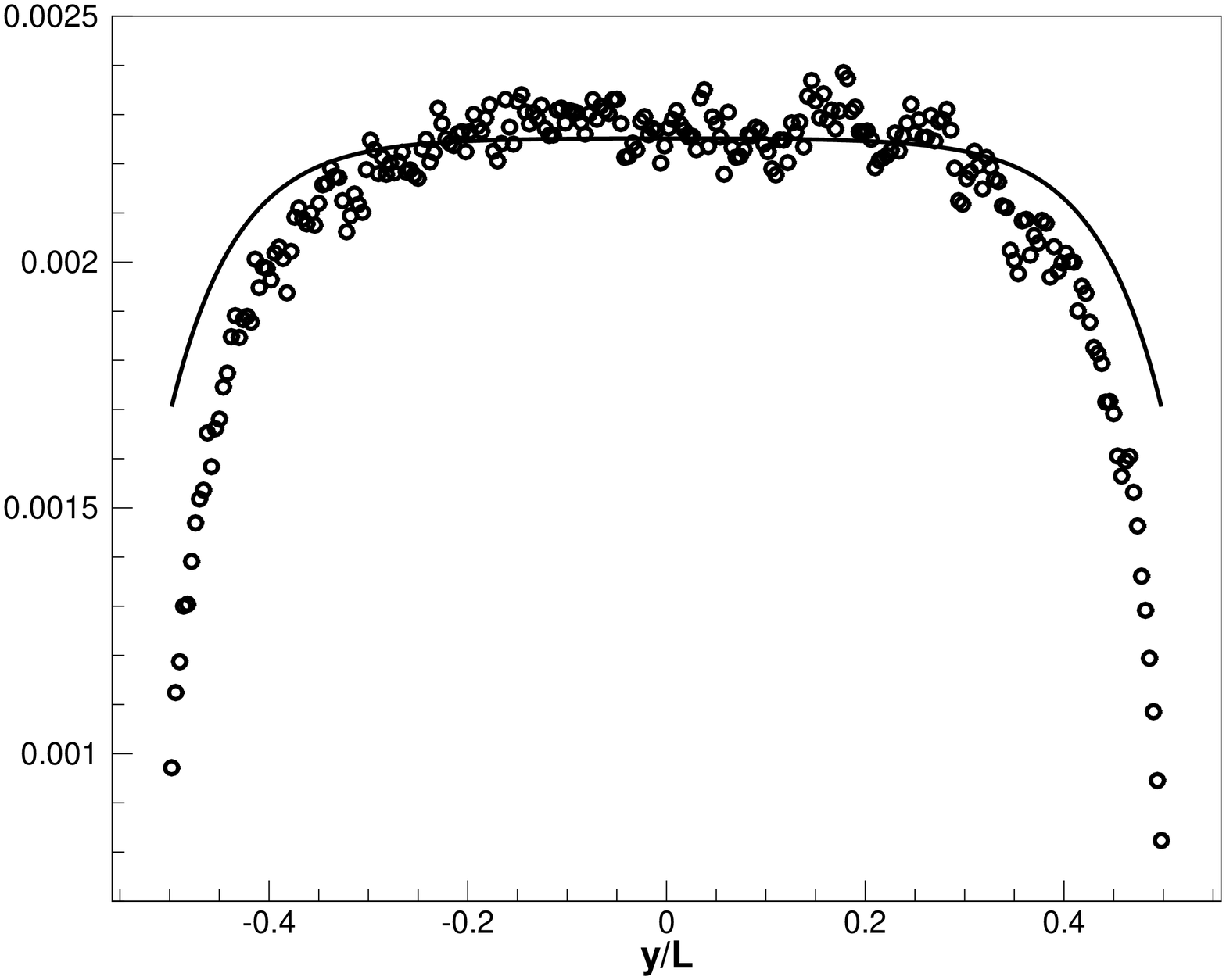,width=0.5\linewidth,clip=}
\end{tabular}
\caption{Left: Shear stress at various Knudsen numbers. Labels are
as in Fig.\ \ref{Fig2}. Plotted is the reduced function
$P^*_{xy}=P_{xy}/P_{xy}^{\infty}$ where $P_{xy}^{\infty}$ is the
shear stress at $\Kn\to\infty$ of the Boltzmann-BGK model
\cite{Willis62}. Right: Nonequilibrium normal stress difference at
$\Kn=0.6$. Line: solution (\ref{NormalStressResult}); Symbol: DSMC
simulation.} \label{Fig3}
\end{figure}

Thirdly and finally, we apply the classical diffuse boundary
conditions \cite{Cerci}, which
were adapted to the present model in Ref. \cite{AK4}. At the
bottom plate ($y= - L/2$), diffuse boundary condition in the
steady state is (see Fig.\ \ref{Fig1}),
%
%
\begin{equation}
f_{2,5,6} \bigl|_{y=-L/2}=f_{2,5,6}^{\rm eq}(\rho,\rho U_1,0).
\label{PopulationsBottom}
\end{equation}
In other words, in the steady state, the diffuse boundary
condition reduces to setting the corresponding populations at
equilibrium (\ref{TQA}) with the density $\rho$ and velocity of
the wall. Now, in order to find a relation between the two
integration constants, $V$ and $\pi$, we notice that the
difference $\left[f_5^{\rm neq} - f_6^{\rm neq} \right]_{y=-L/2}$
can be evaluated in two ways. On one hand, $[f_i^{\rm
neq}]_{y=-L/2}=[f_i]_{y=-L/2}-[f_i^{\rm eq}]_{y=-L/2}$, where the
first contribution is due to (\ref{PopulationsBottom}), and the
second is due to the stationary solution for the equilibrium
(\ref{TQA}) with the velocity (\ref{VelExp}), whereupon
$\left[f_5^{\rm neq} - f_6^{\rm neq} \right]_{y=-L/2}=
\frac{\sqrt{3}{\rho}}{18 c_{\rm s}}(U_1 \!+\! \frac{\pi }{\tau
c_{\rm s}^2}(-\frac{L}{2} \!+\! \tau V))$.
 On other hand, using
(\ref{fneqsol}), we find $\left[f_5^{\rm neq} - f_6^{\rm neq}
\right]_{y=-L/2}
  =\frac{{\rho}\pi}{6c_{\rm s}^2}$.
Matching these expressions, we find a relation between integration
constants $\pi$ and $V$:
\begin{align}
\label{Pxyexp1}
\begin{split}
\pi = \frac{\sqrt{3} \, c_{\rm s}U_1 }{ 3 + \frac{\sqrt{3} }{
\tau\, c_{\rm s}}\, \left(\frac{L}{2} - \tau\, V\right) }.
\end{split}
\end{align}
Similarly, at the top plate ($y=L/2$, see Fig.\ \ref{Fig1}),
$f_{4,7,8} \bigl|_{y=-L/2}=f_{4,7,8}^{\rm eq}(\rho,\rho U_2,0)$.
Again, computing the difference $\left[f_7^{\rm neq} - f_8^{\rm
neq} \right]_{y=L/2}$ in two ways as described above, we find:
\begin{align}
\label{Pxyexp2}
\begin{split}
\pi= -  \frac{\sqrt{3}\, c_{\rm s} \, U_2}{ 3 + \frac{\sqrt{3} }{
\tau\, c_{\rm s}}\,   \left(\frac{L}{2} + \tau\, V \right)}.
\end{split}
\end{align}
Comparing (\ref{Pxyexp1}) and (\ref{Pxyexp2}), we find
coefficients $V$ and $\pi$,
 and,  making use of (\ref{VelExp}), we arrive at the result for
 the velocity profile (\ref{VelocityResult}),
 while the non-equilibrium shear stress $P_{xy}^{\rm neq}=\rho\pi$ reads
 \begin{equation}
\label{pxyAns}
 P_{xy}^{\rm neq}=-\frac{\nu\rho}{\Theta_9} \left(\frac{\Delta
 U}{L}\right),
 \end{equation}
 where $\Theta_9$ is given by (\ref{Slope9}). Note that in the
 $D2Q16$ model the result for the shear stress is obtained by
 replacing $\Theta_9$ with $\Theta_{16}$ (\ref{Slope16}).
 Results for the shear stress are compared with the data of Willis
 \cite{Willis62} and DSMC simulation
 in Fig.\ \ref{Fig3}. In Tab.\ \ref{Limits}, the limiting values of
 the effective shear viscosity, $\nu_{\rm eff}=-P^{\rm
neq}_{xy}L(\Delta U\rho)^{-1}$, at the infinite Knudsen number for
the $D2Q9$ and
 the $D2Q16$ LB models are compared with the Boltzmann-BGK result
 \cite{Willis62}.
\begin{center}
\begin{table}
\begin{tabular}{|l|l|}
  \hline
  $D2Q9$ & $0.723$ \\
  \hline
  $D2Q16$  & $1.113$ \\
  \hline
  BGK \cite {Willis62}& 1 \\
  \hline
\end{tabular}
\caption{Effective shear viscosity $\nu_{\rm eff}$ at
$\Kn\to\infty$, in units of $\sqrt{\frac{k_{\rm B}T_0}{2\pi m}}L$
\cite{Willis62}.} \label{Limits}
\end{table}
\end{center}
Same method is used in order to evaluate the two remaining
integration constants $A$ and $B$. Namely, we evaluate $f_2^{\rm
neq}$ (bottom plate) and $f_4^{\rm neq}$ (top plate) in the two
ways as described above. After some algebra, we find the results
for the non-equilibrium normal stress difference and the
transversal energy flux
\begin{eqnarray}
\normSt^{\rm neq} \!&=&\! \rho\left(\frac{\Delta
U}{L}\right)^2\frac{\tau\nu}{(1+2\Kn)^2}
 \left[
2 \!-\! e^{-\frac{1}{2\Kn}}\cosh{\left(\frac{y}{\Kn L}\right)}
\right],\label{NormalStressResult}\\
q_y^{\rm neq} \!&=&\!-\rho\left(\frac{\Delta U}{L}\right)^2
\frac{\nu}{(1+2\Kn)^2} \!\left[2y \!-\!\Kn L
 e^{-\frac{1}{2\Kn}}
\sinh{\left(\frac{y}{\Kn L}\right)} \right]\!+\!2UP^{\rm
neq}_{xy}.\label{EnergyFluxResult}
\end{eqnarray}

Expressions for the velocity $u_x$ (\ref{VelocityResult}), the
shear stress $P^{\rm neq}_{xy}$ (\ref{pxyAns}), the normal stress
difference $N^{\rm neq}$ (\ref{NormalStressResult}), and the
energy flux $q_y^{\rm neq}$ (\ref{EnergyFluxResult}), when
substituted into $f_i=f_i^{\rm eq}+f_i^{\rm neq}$ (see (\ref{TQA})
and (\ref{fneqsol})), furnish the exact solution of Couette flow
for the nonlinear $D2Q9$ model. The same solution method is
applicable to any member of the LB hierarchy although algebra
becomes more involved. Solution of the $D2Q16$ model leading to
the result (\ref{VelocityResult2}) will be presented in a detailed
publication.

 We have already mentioned that the solution for the velocity
 (\ref{VelocityResult}) is Galilean invariant. Same holds also for the higher-order
 moments. Indeed, the stress tensor is obviously Galilean invariant (that is,
 $P^{\rm neq}_{xy}$ (\ref{pxyAns}) and $N^{\rm neq}$ (\ref{NormalStressResult})
 depend only on $U_2-U_1$),
  and also the energy
 flux transforms correctly under $U_{1,2}'=U_{1,2}+C$:
 from the definition of the energy flux
 $q_y' = q_y+2CP_{xy}+C^2j_y$, which is manifested in
 (\ref{EnergyFluxResult}) ($j_y=0$ in the present solution).
The normal stress difference (\ref{NormalStressResult}) is a
positive-definite function which is consistent with kinetic theory
of gases. Importantly, the nonvanishing of $N^{\rm neq}$ and
$q_y^{\rm neq}$ is the direct implication of the nonlinearity of
the kinetic equation (\ref{LBM}) (manifested by the $\Delta U^2$
dependence in (\ref{NormalStressResult}) and
(\ref{EnergyFluxResult})), and cannot be predicted on the basis of
linearized kinetic theory \cite{Willis62,Cerci}. For that reason,
the DSMC method \cite{Bird} was used in order to validate the
solution in the nonlinear domain. Parameters of the DSMC
simulation correspond to the hard sphere model of Argon at normal
conditions and diffuse scattering at the walls was implemented.
The value of the velocity $U_1=-U_2=0.5C_{\rm s}$, where $C_{\rm
s}=\sqrt{5k_{\rm B}T/3m}$ is the speed of sound of the
three-dimensional ideal gas, was used to maintain a
quasi-isothermal flow. The normal stress difference
(\ref{NormalStressResult}) is mapped onto DSMC data in Fig.\
\ref{Fig3} which qualitatively confirms the prediction.

\begin{center}
\begin{figure}
\includegraphics[scale=0.4]{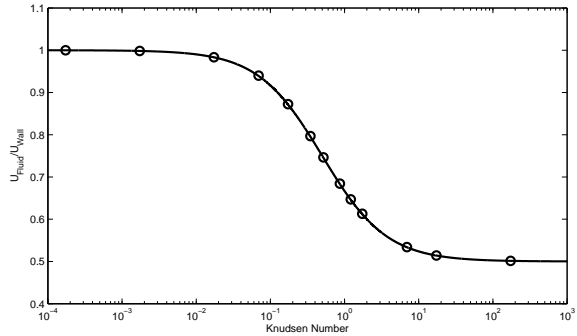}
\caption{Exact solution for the relative slip velocity (line) and
the lattice Boltzmann simulation (symbol) as a function of Knudsen
number.} \label{Fig4}
\end{figure}
\end{center}

Finally, in view of the concerns mentioned in the introduction, we
validate the lattice Boltzmann method for the kinetic equation
(\ref{LBM}). The LB space-time discretization of (\ref{LBM}) is
standard, see, e.\ g.\ \cite{HeChenDoolen98}. Simulation setup
used corresponds to $U_1=0$, periodic boundary conditions were
applied in the $x$-direction. Simulation were run till the steady
state was reached, a grid convergence study was also performed. It
is evident in Fig.\ \ref{Fig4} that the simulated relative slip
accurately reproduces the exact solution (\ref{VelocityResult}).
Same holds for all other moments. Thus, applications of the LBM to
microflows are by no means an ``artifact of numerics"
\cite{LUOCOMMENT}. Importance of physically relevant boundary
conditions must be stressed. We have verified that if the standard
``bounce-back" boundary condition of the LB method is used in the
present problem, then the analytical result predicts vanishing
slip velocity at all Knudsen numbers; thus, results of micro-flow
simulations with the ``bounce-back" boundary conditions or their
derivatives are questionable indeed \cite{Shen03bs}.

To conclude, our analytical results suggest that the hierarchy of
lattice Boltzmann models is the way to approximate the kinetic
theory. Without a denial of a body of LB simulations, it must be
appreciated that only the exact solutions answer unambiguously the
question of the physical validity of the method. Our results
reveal that applications to LB methods to microflows should be
based on LB models with larger velocity sets if one seeks a
quantitative prediction, especially in the transient regime.
 It would be
quite interesting to extend the approach developed herein to
obtain closed-form solutions in other cases of kinetic theory
\cite{Cerci}, and eventually also to three-dimensional cases. In
that respect, we were able to extend the present analysis to the
three-dimensional $D3Q27$ model, results will be reported
elsewhere.

We acknowledge useful discussions of some of the results with G.
Doolen, H.C. \"Ottinger, S. Succi and V. Yakhot. Support by BFE
Project 100862 (I.V.K.), by the ETH Project 0-20235-05 (N.I.P), by
CCEM-CH (I.V.K. and S.A.) and by NTU Office of Research (S.A.) is
gratefully acknowledged.

\bibliography{ExactCouette}

\end{document}